\documentclass[conference]{IEEEtran}

\usepackage{amsfonts}
\usepackage{amssymb}
\usepackage{subfig}
\usepackage{amsmath,dsfont}
\usepackage{amsthm}
\usepackage{mathrsfs}
\usepackage{verbatim}
\usepackage{graphicx}
\usepackage{epstopdf}
\usepackage{mathtools}

\DeclareMathAlphabet{\mathpzc}{OT1}{pzc}{m}{it}

\theoremstyle{remark} \newtheorem{remark}{Remark}

\newcommand{\thmlabel}[1]{\label{thm:#1}}

\theoremstyle{remark} \newtheorem{theorem}{Theorem}

\newcommand{\Xvec}{\mathbf{X}}
\newcommand{\Yvec}{\mathbf{Y}}

\newcommand{\sfQ}{\mathsf{Q}}

\DeclareMathAlphabet{\mathbfsf}{\encodingdefault}{\sfdefault}{bx}{n}

\newcommand{\realSet}{\mathfrak{R}}

\newcommand{\eps}{\epsilon}

\newcommand{\E}{\mathds{E}}
\newcommand{\Ec}[1]{\mathds{E} \left\{ #1 \right\}}

\newcommand{\mN}{\mathcal{N}}

\IEEEoverridecommandlockouts
\title{Energy-Distortion Tradeoff for the Gaussian Broadcast Channel with Feedback
\thanks{
\noindent
This work was supported by Israel Science Foundation under grant 396/11. 
}
}

\author{
Yonathan Murin$^1$, Yonatan Kaspi$^2$, Ron Dabora$^3$, and Deniz G\"und\"uz$^4$ \\
{\small $^1$Stanford University, USA, $^2$University of California, San Diego, USA,} {\small $^3$Ben-Gurion University, Israel, $^4$Imperial College London, UK}
\vspace{-0.6cm}
}
\renewcommand\footnotemark{}
\begin{document}
\maketitle

\begin{abstract}
	This work focuses on the minimum transmission energy required for communicating a pair of correlated Gaussian sources over a two-user Gaussian broadcast channel with noiseless and causal channel output feedback (GBCF).
	We study the fundamental limit on the required transmission energy for broadcasting a pair of source samples, such that each source can be reconstructed at its respective receiver to within a target distortion, when	the source-channel bandwidth ratio is not restricted.
	We derive a lower bound and three distinct upper bounds on the minimum required energy. For the upper bounds we analyze three transmission schemes: Two schemes are based on separate source-channel coding, and apply coding over multiple samples of source pairs. The third scheme is based on joint source-channel coding obtained by extending the Ozarow-Leung (OL) transmission scheme, which applies uncoded linear transmission.
		%
	Numerical simulations show that despite its simplicity, the energy-distortion tradeoff of the OL-based scheme is close to that of the better separation-based scheme, which indicates that the OL scheme is attractive for energy-efficient source transmission over GBCFs.
\end{abstract}


\vspace{-0.23cm}
\section{Introduction}	

\vspace{-0.1cm}
We study the {\em energy-distortion tradeoff} (EDT) for the transmission of a correlated pair of Gaussian sources over a two-user Gaussian broadcast channel (GBC) with noiseless causal feedback, referred to as the GBCF. 
EDT, recently proposed for the multiple-access channel (MAC) and the point-to-point channel in \cite{Jain:12}, characterizes the minimum energy-per-source sample required to achieve a target distortion pair at the receivers, without constraining the source-channel bandwidth ratio.
In many practical scenarios, e.g., satellite broadcasting \cite{AG:11}, sensor networks measuring physical processes \cite{RBC:14}, and in particular wireless body-area sensor networks \cite{NLN:14}, an extremely high energy efficiency is required for broadcasting correlated observations, while the available power is limited.
It follows that there is a strong motivation for studying the EDT for broadcasting correlated sources, and in the present work we focus on EDT for feedback-assisted scenarios, represented by the GBCF.

%

It is well known that, for lossy source transmission over memoryless Gaussian point-to-point channels, with or without feedback, when the bandwidth ratio is fixed and the average power per channel use is limited, separate source and channel coding (SSCC) achieves the minimum possible average distortion \cite[Thm. 3]{Shannon:59}. 
In \cite[Cor. 1]{Jain:12} it is further shown that SSCC is optimal in the EDT sense as well: For any target distortion level, the minimal transmission energy is achieved by optimal lossy compression followed by the most energy efficient channel coding. 

In contrast to the point-to-point setting, in multi-user scenarios joint design of the source and channel codes can improve the performance \cite{Cover:80}. 
However, we have a relatively limited understanding of the impact of feedback on joint source-channel coding (JSCC) over multi-user channels.
The work \cite{Ong:07} presented several achievability schemes and a set of necessary conditions for losslessly transmitting a pair of discrete and memoryless correlated sources over a MAC with feedback. Lossy transmission of correlated Gaussian sources over a two-user Gaussian MAC with feedback was studied in \cite{Lapidoth:10}, in which sufficient conditions as well as necessary conditions for the achievability of a mean square error (MSE) distortion pair were derived. 
While \cite{Lapidoth:10} considered only source-channel coding with a unit bandwidth ratio, \cite{Jain:12} studied the EDT for the transmission of correlated Gaussian sources over a two-user Gaussian MAC with and without feedback, when the bandwidth ratio is not restricted.

Previous works on GBCFs mainly studied the channel coding aspects which applies to independent
and uniformly distributed messages \cite{OzarowBC:84}--\cite{AMM:12}. 
JSCC of correlated Gaussian sources over GBCFs was also studied in \cite{MKDG:Submitted15}, in which the minimal number of channel uses required to achieve a target MSE distortion pair was characterized for three linear encoding schemes, using uncoded transmission: The first scheme is a JSCC scheme based on the coding scheme of Ozarow and Leung \cite{OzarowBC:84}, to which we shall refer as the {\em OL scheme}; the second scheme iss a JSCC scheme based on the linear quadratic Gaussian (LQG) coding scheme of \cite{AMM:12}, to which we shall refer as the {\em LQG scheme}; and the third scheme iss a JSCC scheme whose parameters are obtained using dynamic programming (DP).\footnote{In the present work we discuss only the former two schemes, since the scheme based on DP becomes analytically and computationally untractable as the number of channel uses goes to infinity.}
We note that the advantages of linear and uncoded transmission, as implemented in the OL and LQG schemes, include a low computational complexity, low coding delays, and low storage requirements. 
In the present work we analyze lossy source coding over GBCFs using SSCC and JSCC broadcasting schemes based on a different performance metric -- the EDT.

{\bf{\slshape Main Contributions}:}
This is the {\em first} work towards characterizing the EDT in GBCFs.
We derive lower and upper bounds on the minimum energy required to achieve a target MSE distortion pair when transmitting a pair of Gaussian sources over a two-user GBCF, without constraining the number of channel uses per source sample. 
The proposed lower bound is based on cut-set arguments, and the upper bounds are obtained using three transmission schemes: Two SSCC schemes and an uncoded JSCC scheme.
The first SSCC scheme {\em jointly} compresses the two source sequences into a single bit stream, and transmits this stream using a channel code, designed for the transmission of a common message. 
The second SSCC scheme {\em separately} encodes each source sequence into two distinct bit streams, and then transmits the resulting two bit streams via the LQG channel code of \cite{AMM:12}. 
It is shown that, in terms of the minimum energy per bit, the LQG code provides no gain compared to orthogonal transmission, from which we conclude that the first SSCC scheme, which jointly compresses the sequences into a single stream, is more energy efficient.
As both SSCC schemes apply coding over multiple samples of the source pairs, they require a high computational complexity, long delays, and large storage. Thus, motivated by the low computational complexity, short coding delay, and small storage requirements of the uncoded OL scheme presented in \cite{MKDG:Submitted15}, we use this scheme to obtain a third upper bound on the EDT.

%

Numerical results indicate that the SSCC scheme based on joint compression achieves a better EDT compared to the JSCC OL scheme; yet, in many scenarios the gap is quite small. 
Moreover, in many applications there is a constraint on the maximal allowed latency.
In such scenarios, coding over large blocks of independent and identically distributed (i.i.d.) pairs of source samples introduces unacceptable delays, and instantaneous transmission of each observed pair of source samples via the proposed JSCC-OL scheme may be preferable in order to satisfy the latency requirement, while maintaining high energy efficiency.

		
The rest of this paper is organized as follows: The problem formulation is detailed in Section \ref{sec:ProbForm}. Bounds on the minimum energy are derived in Section \ref{sec:LowerBound}. 
Numerical results are described in Section \ref{sec:numerical}, and concluding remarks are provided in Section~\ref{sec:conclusions}.

{\bf{\slshape Notation}:} We use capital letters to denote random variables, e.g., $X$, and boldface letters to denote column random vectors, e.g., $\Xvec$, where the $k$'th element of a vector $\Xvec$ is denoted by $X_k, k \mspace{-3mu} \ge \mspace{-3mu} 1$.
We use sans-serif fonts to denote matrices, e.g., $\sfQ$.
Finally, we use $\Ec{\cdot}$, $(\cdot)^T$, $\log (\cdot)$, and $\realSet$ to denote stochastic expectation, transpose, natural basis logarithm, and the set of real numbers, respectively.

\vspace{-0.15cm}
\section{Problem Definition} \label{sec:ProbForm}

\begin{figure}[t]
    \centering
    \includegraphics[width=0.7\columnwidth]{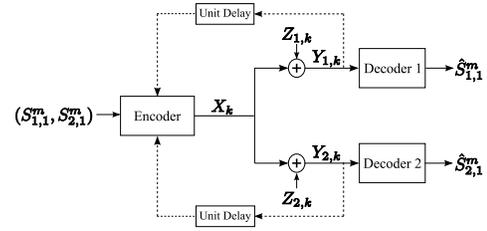}
    \captionsetup{font=footnotesize}
		\caption{Gaussian broadcast channel with correlated sources and feedback links. $\hat{S}_{1,1}^m$ and $\hat{S}_{2,1}^m$ are the reconstructions of $S_{1,1}^m$ and $S_{2,1}^m$, respectively.}
    \label{fig:GBC}
   	\vspace{-0.45cm}
\end{figure}

\vspace{-0.1cm}
Fig. \ref{fig:GBC} depicts the two-user GBCF, in which all the signals are real. 
The encoder observes $m$ i.i.d. realizations of a correlated and jointly Gaussian pair of sources $(S_{1,j},S_{2,j}) \mspace{-4mu} \sim \mspace{-4mu} \mN(0, \sfQ_s),j \mspace{-3mu}= \mspace{-3mu}1,\dots,m$, where $\sfQ_s \mspace{-3mu} \triangleq \mspace{-3mu} \sigma_s^2 \mspace{-2mu} \cdot \mspace{-2mu} \left[ \begin{smallmatrix}1 & \rho_s \\ \rho_s & 1 \end{smallmatrix} \right]$, $|\rho_s| \mspace{-3mu} < \mspace{-3mu} 1$. 
Both sources generate their realizations at a fixed rate. The task of the encoder is to send the observations of the $i$'th source $S_{i,1}^m, i \mspace{-3mu} = \mspace{-3mu} 1,2$, to the $i$'th decoder (receiver) denoted by Rx$_i$.
%
The received signal at time~$k$~at~Rx$_i$~is~given~by:
\vspace{-0.2cm}
\begin{align}
	Y_{i,k} & = X_k + Z_{i,k}, \quad i=1,2, \label{eq:signalModel}
\end{align}

\vspace{-0.15cm}
\noindent for $k \mspace{-3mu} = \mspace{-3mu} 1,\dots,n$, where the noise sequences $\{Z_{1,k}, Z_{2,k}\}_{k=1}^n$, are i.i.d. over $k \mspace{-3mu} = \mspace{-3mu} 1,2,\dots,n$, with $(Z_{1,k},Z_{2,k}) \mspace{-4mu} \sim \mspace{-4mu} \mN(0, \sfQ_z)$, where $\sfQ_z \mspace{-3mu} \triangleq \mspace{-3mu} \sigma_z^2 \mspace{-2mu} \cdot \mspace{-2mu} \left[ \begin{smallmatrix}1 & \rho_z \\ \rho_z & 1 \end{smallmatrix} \right]$, $|\rho_z| \mspace{-3mu} < \mspace{-3mu} 1$.
Let $\Yvec_{k} \mspace{-3mu} \triangleq \mspace{-3mu} (Y_{1,k}, Y_{2,k})$.
%
%
Rx$_i$, $i \mspace{-3mu} = \mspace{-3mu} 1,2$, uses its channel output sequence $Y_{i,1}^n$ to estimate $S_{i,1}^m$ via $\hat{S}_{i,1}^{m} \mspace{-5mu} = \mspace{-5mu} g_{i}(Y_{i,1}^n), g_{i} \mspace{-5mu} : \mspace{-5mu} \realSet^{n} \mspace{-5mu} \rightarrow \mspace{-5mu} \realSet^m$. The encoder maps the observed pair of source sequences and the noiseless causal channel outputs obtained through the feedback links into a channel input via: $X_k \mspace{-3mu} = \mspace{-3mu} f_k(S_{1,1}^m,S_{2,1}^m,\Yvec_{1}, \Yvec_{2}, \dots, \Yvec_{k-1}), f_k \mspace{-3mu} : \mspace{-3mu} \realSet^{2(m+k-1)} \mspace{-3mu} \rightarrow \mspace{-3mu} \realSet$.

We study the symmetric GBCF with parameters $(\sigma_s^2, \rho_s, \sigma_{z}^2, \rho_z)$, and define a $(D,E,m,n)$ code to be a collection of $n$ encoding functions $\{f_k \}_{k=1}^n$ and two decoding functions $g_1,g_2$, that satisfy the MSE distortion constraint:
\vspace{-0.2cm}
\begin{align}
	\sum_{j=1}^m \E \{ (S_{i,j} \mspace{-3mu} - \mspace{-3mu} \hat{S}_{i,j})^2 \} \le m D, \quad 0 \mspace{-3mu} < \mspace{-3mu} D \mspace{-3mu} \le \mspace{-3mu} \sigma_s^2, \quad i=1,2,
	\label{eq:distortionDef}
\end{align}

\vspace{-0.1cm}
\noindent and energy constraint:
\vspace{-0.15cm}
\begin{align}
	\sum_{k=1}^n \Ec { X_k^2 } \le m E. \label{eq:EnergyConstraint}
\end{align}

\vspace{-0.15cm}
\noindent Constraint \eqref{eq:EnergyConstraint} reflects the energy per source sample rather than per channel use. Note that by defining $P \triangleq \frac{m}{n} E$, constraint \eqref{eq:EnergyConstraint} can be equivalently stated as $ \frac{1}{n} \sum_{k=1}^n \Ec { X_k^2 } \le P$, which is the well known average power constraint.
Our objective is to characterize the minimal $E$, for a given target MSE $D$ at each user, such that for all $\eps > 0$ there exists $m,n$ and a $(D+\eps,E + \eps,m,n)$ code. 
We call this minimal value the EDT, and denote it by $E(D)$. In the next section we present lower and upper bounds on $E(D)$.

\vspace{-0.07cm}
\section{Bounds on $E(D)$} \label{sec:LowerBound}

\vspace{-0.1cm}
\subsection{Lower Bound on $E(D)$}

\vspace{-0.1cm}
Our first result is a lower bound on $E(D)$.
We begin by defining the following rate-distortion functions, see \cite[Sec. III.B]{Lapidoth:10B}:
\begin{subequations} \label{eq:rateDistFuncs}
\begin{align}
	\mspace{-8mu} R_{S_1}(D) & \mspace{-4mu} \triangleq \mspace{-4mu} \frac{1}{2} \log_2 \left( \frac{\sigma_s^2}{D} \right) \label{eq:Rs1(D)} \\
	\mspace{-8mu} R_{S_1,S_2}(D) & \mspace{-4mu} \triangleq \mspace{-4mu} \begin{cases} \mspace{-3mu} \frac{1}{2} \mspace{-2mu} \log_2 \mspace{-2mu} \left( \mspace{-2mu} \frac{\sigma_s^2(1+|\rho_s|)}{2D - \sigma_s^2(1-|\rho_s|)} \mspace{-2mu} \right) \mspace{-2mu} , & \mspace{-7mu} D \mspace{-4mu} > \mspace{-4mu} \sigma_s^2(1 \mspace{-4mu} - \mspace{-4mu} |\rho_s|) \\ \mspace{-2mu} \frac{1}{2} \mspace{-2mu} \log_2 \mspace{-2mu} \left( \mspace{-2mu} \frac{\sigma_s^4(1-\rho_s^2)}{D^2} \mspace{-2mu} \right) \mspace{-2mu} , & \mspace{-7mu} D \mspace{-4mu} \le \mspace{-4mu} \sigma_s^2(1 \mspace{-4mu} - \mspace{-4mu} |\rho_s|) \end{cases} \mspace{-3mu} . \label{eq:Rs1s2(D)}
\end{align}
\end{subequations}

\noindent Note that $R_{S_1}(D)$ is the rate-distortion function for the source variable $S_1$, while $R_{S_1,S_2}(D)$ is the rate distortion function for jointly compressing the pair of sources $\{S_1,S_2\}$ into a single stream of rate $R \ge R_{S_1,S_2}(D)$. Note that \cite[Sec. III.B]{Lapidoth:10B} uses the function $R_{S_1,S_2}(D_1,D_2)$ as it considers a different distortion constraint for each source. 
For the present case, in which the same distortion constraint is applied to both sources, $R_{S_1,S_2}(D)$ can be obtained by setting $D_1=D_2=D$ in \cite[Eq. (10)]{Lapidoth:10B}, and thus we use the simplified notation $R_{S_1,S_2}(D)$.
The lower bound on $E(D)$ is stated in the following theorem:
\begin{theorem} \thmlabel{thm:lowerBound}
	The EDT $E(D)$ satisfies $ E(D) \mspace{-3mu} \ge \mspace{-3mu} E_{\text{lb}}(D)$, where:
	\begin{align}
    \mspace{-8mu}  E_{\text{lb}}( \mspace{-1mu} D \mspace{-1mu}) \mspace{-3mu} = \mspace{-3mu} \sigma_z^2 \mspace{-2mu} \cdot \mspace{-2mu} \log_e 2 \mspace{-2mu} \cdot \mspace{-2mu} \max \mspace{-2mu} \big\{ 2R_{S_1}(D), (1 \mspace{-3mu} + \mspace{-3mu} \rho_z) R_{S_1,S_2}(D) \big\} \mspace{-1mu}.
		\label{eq:EDlowerBound}
	\end{align}
\end{theorem}

\begin{IEEEproof}[$\mspace{-35mu}$ Proof Outline]  
	As we consider a symmetric setting, we can focus on the distortion at Rx$_1$.	
	To obtain \eqref{eq:EDlowerBound}, we derive two different lower bounds. 
	The first bound is derived by ignoring Rx$_2$ and requiring that the distortion at Rx$_1$ will be $D$.
	The second bound is obtained by considering the transmission of both sources over a point-to-point channel with two outputs $Y_1$ and $Y_2$. 
	In \cite[Sec. III]{MKDG_EDT:15} we explicitly compute the resulting rate bounds, and show that, if a $(D \mspace{-3mu} + \mspace{-3mu} \eps, E \mspace{-3mu} + \mspace{-3mu} \eps, m ,n), \epsilon \mspace{-3mu} > \mspace{-3mu} 0$, code exists, then the first bound yields $R_{S_1}(D) \mspace{-3mu} \le \mspace{-3mu} \frac{ (E + \eps)}{2 \sigma_z^2 \log_e 2}$, and the second bound yields $R_{S_1,S_2}(D) \mspace{-3mu} \le \mspace{-3mu} \frac{(E + \eps)}{\sigma_z^2 (1 + \rho_z) \log_e 2}$. 
\end{IEEEproof}

In the next subsections we analyze the EDT for three achievability schemes thereby obtaining three upper bounds on $E(D)$. 
While the considered schemes have simple code constructions, analyzing their EDT performance analysis is challenging.

\subsection{Upper Bounds on $E(D)$ via SSCC} \label{sec:sepUB}

SSCC in multi-user scenarios carries the advantages of modularity and ease of integration with the layered network architecture.
In this subsection we analyze the EDT of two SSCC schemes: The first scheme takes advantage of the correlation between the sources and ignores the correlation between the noise components. The second scheme ignores the correlation between the sources and aims at utilizing the correlation between the noise components.


\subsubsection{The SSCC-$\rho_s$ Scheme (Utilizing $\rho_s$)} \label{subsec:SSCC_1}

This scheme utilizes the correlation between the sources by first {\em jointly encoding the two source sequences} into a {\em single bit stream} via the source coding scheme proposed in \cite[ Thm. III.1]{Lapidoth:10B}. This step gives rise to the rate-distortion function stated in (4b).
The resulting bit stream is then encoded via an optimal channel code for sending a common message over the GBC (without feedback), and is transmitted to both receivers.
Note that the optimal code for transmitting a {\em common message} over GBCFs with $\rho_z \neq 0$ is not known. When $\rho_z = 0$, the capacity for sending a common message over the GBCF  
is achievable using an optimal point-to-point channel code which ignores the feedback.
Thus, SSCC-$\rho_s$ uses the correlation between the sources, but ignores the correlation among the noise components. 
			The following theorem states the EDT achieved by this scheme.
\begin{theorem} \thmlabel{thm:sepSchemeI}
	The SSCC-$\rho_s$ scheme achieves the following EDT:
	\vspace{-0.1cm}
	\begin{align}
		\mspace{-10mu} E_{\text{sep}}^{(\rho_s)}(D) \mspace{-4mu} = \mspace{-4mu} \begin{cases} \mspace{-2mu} \sigma_z^2 \log_e \mspace{-3mu} \left( \mspace{-2mu} \frac{\sigma_s^2(1+|\rho_s|)}{2D - \sigma_s^2(1-|\rho_s|)} \right), & \mspace{-5mu} D \mspace{-3mu} > \mspace{-3mu} \sigma_s^2(1 \mspace{-3mu} - \mspace{-3mu} |\rho_s|) \\ \mspace{-2mu} \sigma_z^2 \log_e \mspace{-3mu} \left( \mspace{-2mu} \frac{\sigma_s^4(1-\rho_s^2)}{D^2} \right), & \mspace{-5mu} D \mspace{-3mu} \le \mspace{-3mu} \sigma_s^2(1 \mspace{-3mu} - \mspace{-3mu} |\rho_s|) \end{cases} \mspace{-3mu} .
		\label{eq:ED_sep1}
	\end{align}
\end{theorem}

\begin{IEEEproof}[$\mspace{-35mu}$ Proof]
	The optimal rate for jointly encoding the source sequences into a single bit stream is $R_{S_1, S_2}(D)$, given in \eqref{eq:Rs1s2(D)} \cite[Sec. III.B]{Lapidoth:10B}.
	Note that from this stream both source sequences can be recovered to within a distortion $D$. 
	The encoded bit stream is then transmitted to the receivers via a capacity-achieving point-to-point channel code \cite[Thm. 10.1.1]{cover-thomas:it-book} (note that this code does not need the causal feedback \cite[Thm. 8.12.1]{cover-thomas:it-book}).
	Let $E_{b \min}^{\text{common}}$ denote the minimum energy-per-bit required for reliable transmission over the Gaussian point-to-point channel \cite{Verdu:90}.
	From \cite[pg.~1025]{Verdu:90}  we have $E_{b \min}^{\text{common}} \mspace{-3mu} = \mspace{-3mu} 2 \sigma_z^2 \log_e 2$.
	As the considered scheme is based on source-channel separation, the achievable EDT is given by $E(D) \mspace{-3mu} = \mspace{-3mu} E_{b \min}^{\text{common}} \times R_{S_1, S_2}(D)$, where $R_{S_1, S_2}(D)$ is stated in \eqref{eq:Rs1s2(D)}. This results in the EDT in \eqref{eq:ED_sep1}.
\end{IEEEproof}


\vspace{0.06cm}
\subsubsection{The SSCC-$\rho_z$ Scheme (Utilizing $\rho_z$)}

This scheme utilizes the correlation among the noise components, which is available through the feedback links for channel encoding, but does not use the correlation between the sources in the compression. 
First, each of the source sequences is encoded using the optimal rate-distortion source code for scalar Gaussian sources \cite[Thm. 13.3.2]{cover-thomas:it-book}. 
Then, the resulting two bit streams are sent over the GBCF using the LQG channel coding scheme of \cite{AMM:12}.
The following theorem characterizes the minimum energy per source sample required by this scheme.

%

\begin{theorem} \thmlabel{thm:sepScheme}
	The SSCC-$\rho_z$ scheme achieves the EDT: $E_{\text{sep}}^{(\rho_z)} \mspace{-2mu} ( \mspace{-1mu} D \mspace{-1mu}) \mspace{-3mu} = \mspace{-3mu} 2\sigma_z^2 \log_e \mspace{-2mu} \left( \mspace{-2mu} \frac{\sigma_s^2}{D} \mspace{-2mu} \right)$.
\end{theorem}

\begin{IEEEproof}
	The encoder separately compresses each source sequence at rate $R_{S_1}(D)$, where $R_{S_1}(D)$ is given in \eqref{eq:Rs1(D)}. 
	Thus, from each encoded stream the corresponding source sequence can be recovered to within a distortion $D$.
	Then, the two encoded bit streams are transmitted to the receivers using the LQG scheme of \cite{AMM:12}. 
	Let $E_{b \min}^{\text{LQG}}$ denote the minimum required energy {\em per pair of encoded bits}, required by the LQG scheme. 	
	In \cite[Appendix. B]{MKDG_EDT:15} we show that for the symmetric scenario $E_{b \min}^{\text{LQG}} \mspace{-3mu} = \mspace{-3mu} 2 \sigma_z^2 \log_e 2$.
	Since two bit streams are transmitted, the achievable EDT is given by $E(D) \mspace{-3mu} = \mspace{-3mu} E_{b \min}^{\text{LQG}} \times 2R_{S_1}(D)$, yielding $E_{\text{sep}}^{(\rho_z)}(D) = 2\sigma_z^2 \log_e \left( \frac{\sigma_s^2}{D} \right)$.
	%
	\end{IEEEproof}

\begin{remark}
Since $E_{\text{sep}}^{(\rho_z)}(D)$ is independent of $\rho_z$, the LQG scheme cannot take advantage of the correlation among the noise components to improve the minimum energy per source sample needed in the symmetric setting. Indeed, an EDT of $E_{\text{sep}}^{(\rho_z)}(D)$ can also be achieved by transmitting the two bit streams via time sharing over the GBCF without using the feedback. 
\end{remark}

\begin{remark} \label{rem:gap_sep_lb}
	We observe that $E_{\text{sep}}^{(\rho_s)}(D) \mspace{-3mu} \le \mspace{-3mu} E_{\text{sep}}^{(\rho_z)}(D)$.
	For $D \mspace{-3mu} \le \mspace{-3mu} \sigma_s^2(1-|\rho_s|)$ this relationship directly follows from the expressions of $E_{\text{sep}}^{(\rho_s)}(D)$ and $E_{\text{sep}}^{(\rho_z)}(D)$. 
	For $D \mspace{-3mu} > \mspace{-3mu} \sigma_s^2(1-|\rho_s|)$ the above relationship holds if the polynomial $q(D) = D^2(1 \mspace{-3mu} + \mspace{-3mu} |\rho_s|) \mspace{-3mu} - \mspace{-3mu} 2\sigma_s^2D \mspace{-3mu} + \mspace{-3mu} \sigma_s^4(1 \mspace{-3mu} - \mspace{-3mu} |\rho_s|)$ is positive. This is satisfied as the the discriminant of $q(D)$ is negative.
	We thus conclude that {\em it is preferable to use the correlation between the sources than the correlation between the noise components}. 				
\end{remark}


\vspace{-0.1cm}
\subsection{An Upper Bound on $E(D)$ via JSCC} \label{subsec:OL}

\begin{figure*}[t!]
	\normalsize
	\captionsetup{font=footnotesize}
	\centering
	\begin{minipage}{.425\textwidth}
		\centering
		\includegraphics[width=1\linewidth]{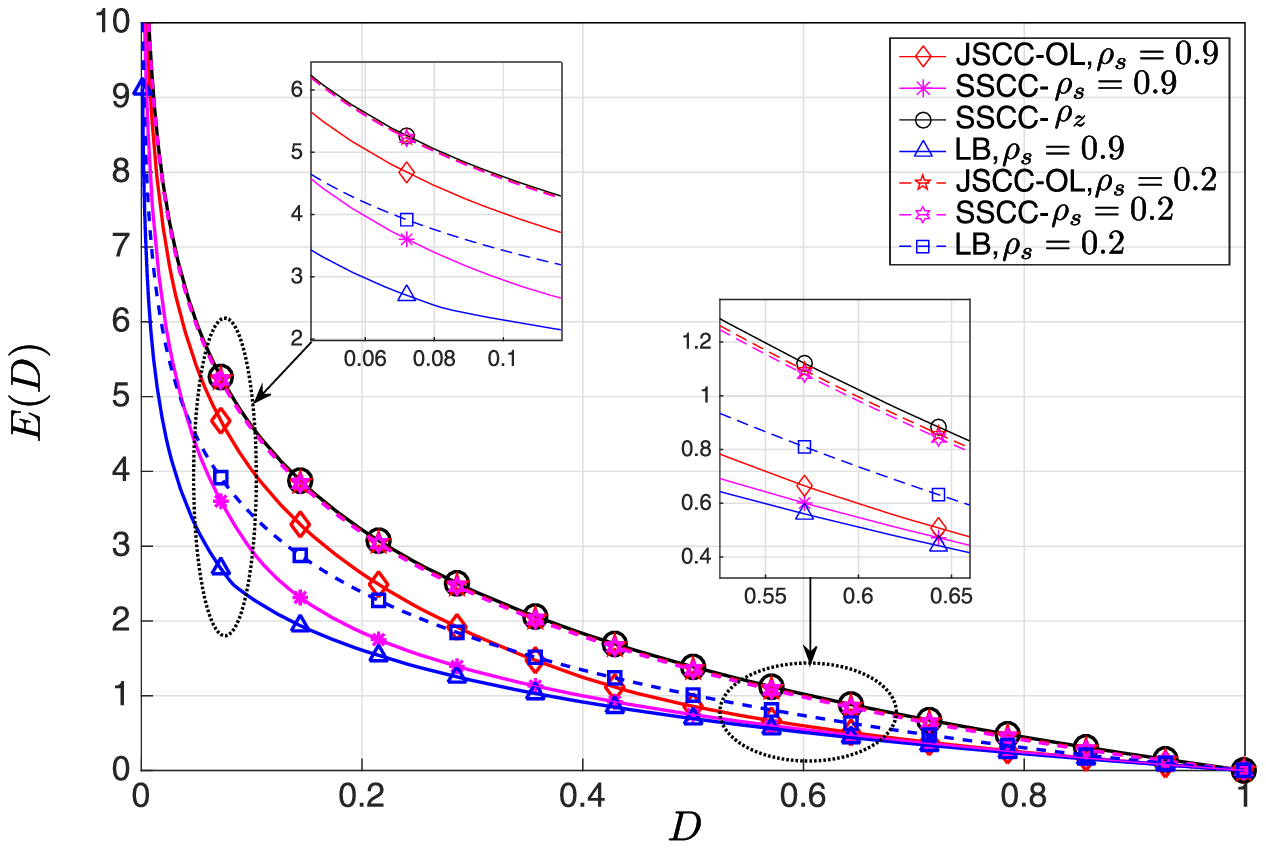}
		\vspace{-0.55cm}
		\captionof{figure}{Upper and lower bounds on $E(D)$ for $\sigma_s^2 \mspace{-3mu} = \mspace{-3mu} \sigma_z^2 \mspace{-3mu} = \mspace{-3mu} 1$, and $\rho_z \mspace{-3mu} = \mspace{-3mu} 0.5$. Solid lines correspond to $\rho_s \mspace{-3mu} = \mspace{-3mu} 0.9$, while dashed lines correspond to $\rho_s \mspace{-3mu} = \mspace{-3mu} 0.2$.}
		\label{fig:ED_fixed_rz}
		\vspace{-0.4cm}
	\end{minipage}
	\hspace{1.5cm}
	\begin{minipage}{.425\textwidth}
		\centering
		\includegraphics[width=1\linewidth]{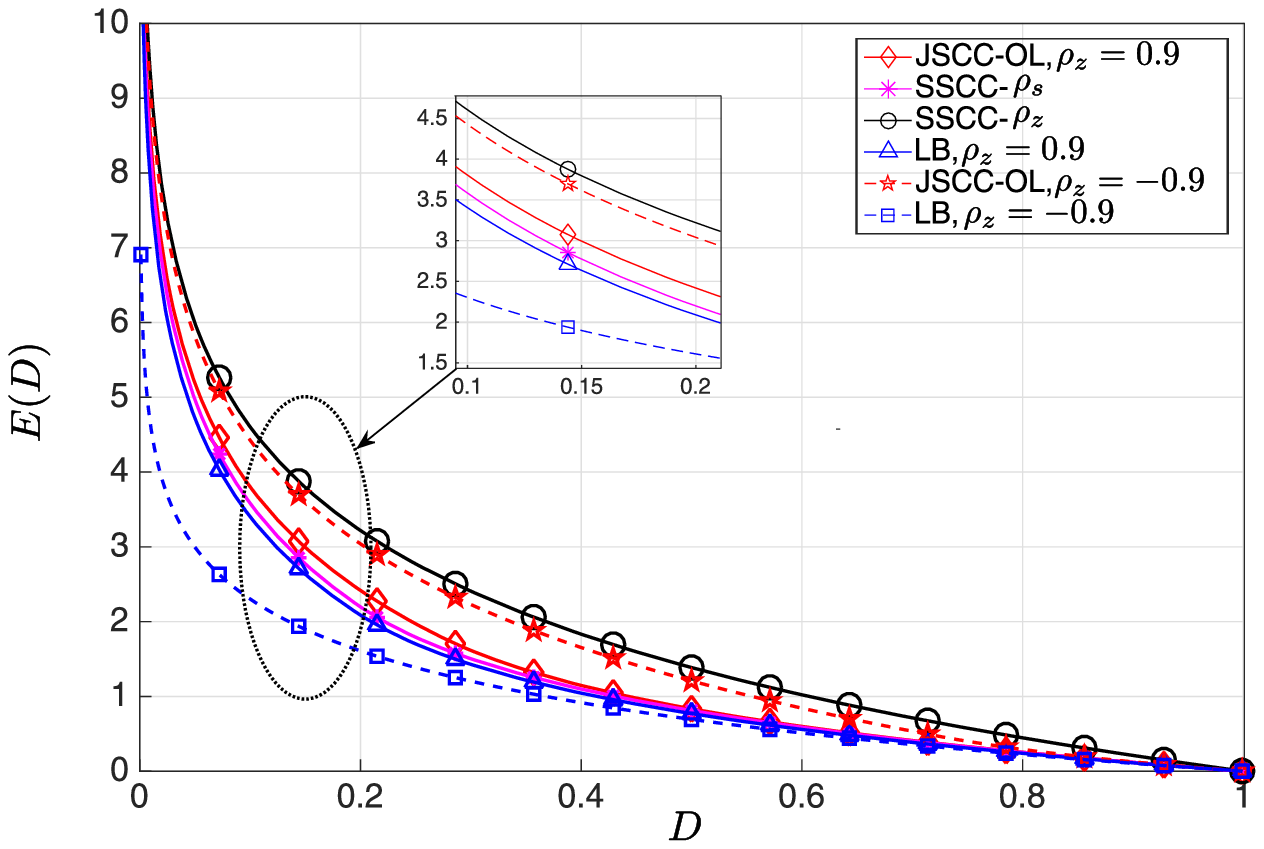}
		\vspace{-0.55cm}
		\captionof{figure}{Upper and lower bounds on $E(D)$ for $\sigma_s^2 \mspace{-3mu} = \mspace{-3mu} \sigma_z^2 \mspace{-3mu} = \mspace{-3mu} 1, \rho_s \mspace{-3mu} = \mspace{-3mu} 0.8$. Solid lines correspond to $\rho_z \mspace{-3mu} = \mspace{-3mu} 0.9$, while dashed lines correspond to $\rho_z \mspace{-3mu} = \mspace{-3mu} -0.9$.}
		\label{fig:ED_fixed_rs}
		\vspace{-0.4cm}
	\end{minipage}
\end{figure*}

\vspace{-0.1cm}
Next, we derive an upper bound on $E(D)$ using the uncoded JSCC OL scheme. This scheme {\em sequentially} transmits the source pairs $(S_{1,j},S_{2,j}), j=1,2,\dots,m$, without applying a source code. 
We note that the OL scheme is designed for a fixed $P \mspace{-3mu} = \mspace{-3mu} \frac{E}{n}$, and from condition \eqref{eq:EnergyConstraint} we obtain that $P \mspace{-3mu} = \mspace{-3mu} \frac{E}{n} \mspace{-3mu} \ge \mspace{-3mu} \frac{1}{n}\sum_{k=1}^n \Ec { X_k^2 }$. An upper bound on $E(D)$ is obtained by calculating the minimal number of channel uses required by the OL scheme to achieve the target distortion $D$, denoted by $K_{\text{OL}}(P,D)$, and then obtaining the required energy via $\sum_{k=1}^{K_{\text{OL}}(P,D)} \Ec { X_k^2 }$. 

In the OL scheme, each receiver recursively estimates its intended source. Using the feedback, the transmitter tracks the estimation errors at the receivers, and sends a linear combination of these errors, where the scheme is terminated after $K_{\text{OL}}(P,D)$ channel uses.
For a detailed description of the proposed OL-based JSCC scheme we refer the reader to \cite[Sec. V.A]{MKDG_EDT:15}.
Let $E_{\text{OL-min}}(D)$ denote the minimal energy per source-pair sample, required to achieve an MSE of $D$ {\em using the JSCC OL scheme}. 
The following theorem presents an upper bound on $E_{\text{OL-min}}(D)$, and therefore on  $E(D)$.
\begin{theorem}
\thmlabel{thm:OL_E(D)_UB}
Let $D_{\text{th}} \mspace{-3mu} \triangleq \mspace{-3mu} \frac{\sigma_s^2 (2 - \rho_z - |\rho_s|)}{2 - \rho_z}$. Then, $E_{\text{OL-min}}(D) \mspace{-3mu} \le \mspace{-3mu} E_{\text{OL}}(D)$, where:
\begin{align} \label{eq:OL_EnegyDistortionTradeoff}
	\mspace{-11mu} E_{\text{OL}} \mspace{-2mu} ( \mspace{-1mu} D \mspace{-1mu} ) \mspace{-4mu} = \mspace{-5mu}
		\begin{cases} \mspace{-4mu} \frac{2 \sigma_z^2}{3 - \rho_z} \log \mspace{-3mu} \left( \mspace{-3mu} \frac{\sigma_s^2(1+|\rho_s|)}{D+(2-\rho_z)(D-\sigma_s^2) + \sigma_s^2 \cdot |\rho_s|} \mspace{-2mu} \right) \mspace{-3mu} , & \mspace{-3mu}  D \mspace{-4mu} \ge \mspace{-4mu} D_{\text{th}} ,\\
		\mspace{-4mu} 2 \sigma_z^2 \Big( \log \mspace{-3mu} \left( \frac{(2- \rho_z - |\rho_s|)\sigma_s^2}{(2- \rho_z)D} \right) \\
		\mspace{20mu} + \frac{1}{3-\rho_z} \log \left( \frac{(2-\rho_z)(1+ |\rho_s|)}{2-\rho_z - |\rho_s|} \right) \mspace{-5mu} \Big), & \mspace{-3mu} D \mspace{-4mu} < \mspace{-4mu} D_{\text{th}}.
		\end{cases}
\end{align}
\end{theorem} 

\vspace{-0.05cm}	
\begin{IEEEproof}[$\mspace{-35mu}$ Proof Outline]
	In \cite[Sec. V.B]{MKDG_EDT:15} we derive an upper bound on $K_{\text{OL}}(P,D)$, denoted by $K_{\text{OL}}^{\text{ub}}(P,D)$. 
	We further show that $\frac{K_{\text{OL}}^{\text{ub}}(P,D)}{K_{\text{OL}}(P,D)} \mspace{-3mu} \to \mspace{-3mu} 1$ and that $K_{\text{OL}}^{\text{ub}}(P,D) \mspace{-3mu} \to \mspace{-3mu} \infty$ as $P \mspace{-3mu} \to \mspace{-3mu} 0$. 
	As both sources generate their realizations at a fixed rate, the above limit ratio implies that the bandwidth used by the OL scheme increases to infinity as $P \mspace{-3mu} \to \mspace{-3mu} 0$.
	Next, recalling that $E_{\text{OL}}(D) \mspace{-3mu} = \mspace{-3mu} P \mspace{-2mu} \cdot \mspace{-2mu} K_{\text{OL}}(P,D)$, we evaluate $P\cdot K_{\text{OL}}^{\text{ub}}(P,D)$ as $P \mspace{-3mu} \to \mspace{-3mu} 0$, and obtain an upper bound on $E_{\text{OL-min}}(D)$. 
	The detailed proof is provided in \cite[Sec. V.C]{MKDG_EDT:15}.
\end{IEEEproof}
	

\begin{remark} \label{rem:OL_discussion}
	It can be observed that when the sources are independent, i.e., $\rho_s \mspace{-3mu} = \mspace{-3mu} 0$, then $E_{\text{OL}}(D) \mspace{-3mu} = \mspace{-3mu} E_{\text{sep}}^{(\rho_s)}(D) \mspace{-3mu} = \mspace{-3mu} E_{\text{sep}}^{(\rho_z)}(D)$, for all $0 \le D \le \sigma_s^2$. 
	When $|\rho_s| \mspace{-2mu} \to \mspace{-2mu} 1$ and $\rho_z \mspace{-2mu} \to \mspace{-2mu} 1$ then $E_{\text{OL}}(D) \mspace{-2mu} \to \mspace{-2mu} E_{\text{lb}}(D) \approx \sigma_z^2 \log_e \left(\frac{\sigma_s^2}{D} \right)$, in this case we also have $E_{\text{sep}}^{(\rho_s)}(D) \mspace{-2mu} \to \mspace{-2mu} E_{\text{lb}}(D)$ and $E_{\text{sep}}^{(\rho_z)}(D) \approx 2 E_{\text{OL}}(D)$. 
\end{remark}

\begin{remark} \label{rem:LQG}
	In this work we did not analyze the EDT of JSCC using the LQG scheme, $E_{\text{LQG}}(D)$. 
	The reason is two-fold: analytic tractability and practical relevance. We note that in \cite[Sec. 4]{MKDG:Submitted15} we adapted the LQG scheme for \cite{AMM:12} to the transmission of correlated Gaussian sources over GBCFs. We have observed in \cite{MKDG:Submitted15} that obtaining a closed-form expression for $E_{\text{LQG}}(D)$ seems intractable. Yet, using the results and analysis of \cite{MKDG:Submitted15} one can find good approximations for $E_{\text{LQG}}(D)$.  
	We also showed in \cite{MKDG:Submitted15} that, in the context of JSCC, and in contrast to the results of \cite{ASW:14} for the channel coding problem, when the duration of transmission is finite and the transmission power is very low, the OL scheme outperforms the LQG scheme. This conclusion is expected to hold for the EDT as well.
Indeed, numerical simulations indicate that the LQG scheme of \cite[Sec. 4]{MKDG:Submitted15} achieves roughly the same minimum energy as the SSCC-$\rho_z$ scheme, while in Section \ref{sec:numerical} we show that the OL scheme outperforms the SSCC-$\rho_z$ scheme.
\end{remark}

\vspace{-0.2cm}
\section{Numerical Results} \label{sec:numerical}

\begin{figure*}[t!]
	\normalsize
	\captionsetup{font=footnotesize}
	\centering
	\begin{minipage}{.41\textwidth}
		\centering
		\includegraphics[width=1\linewidth]{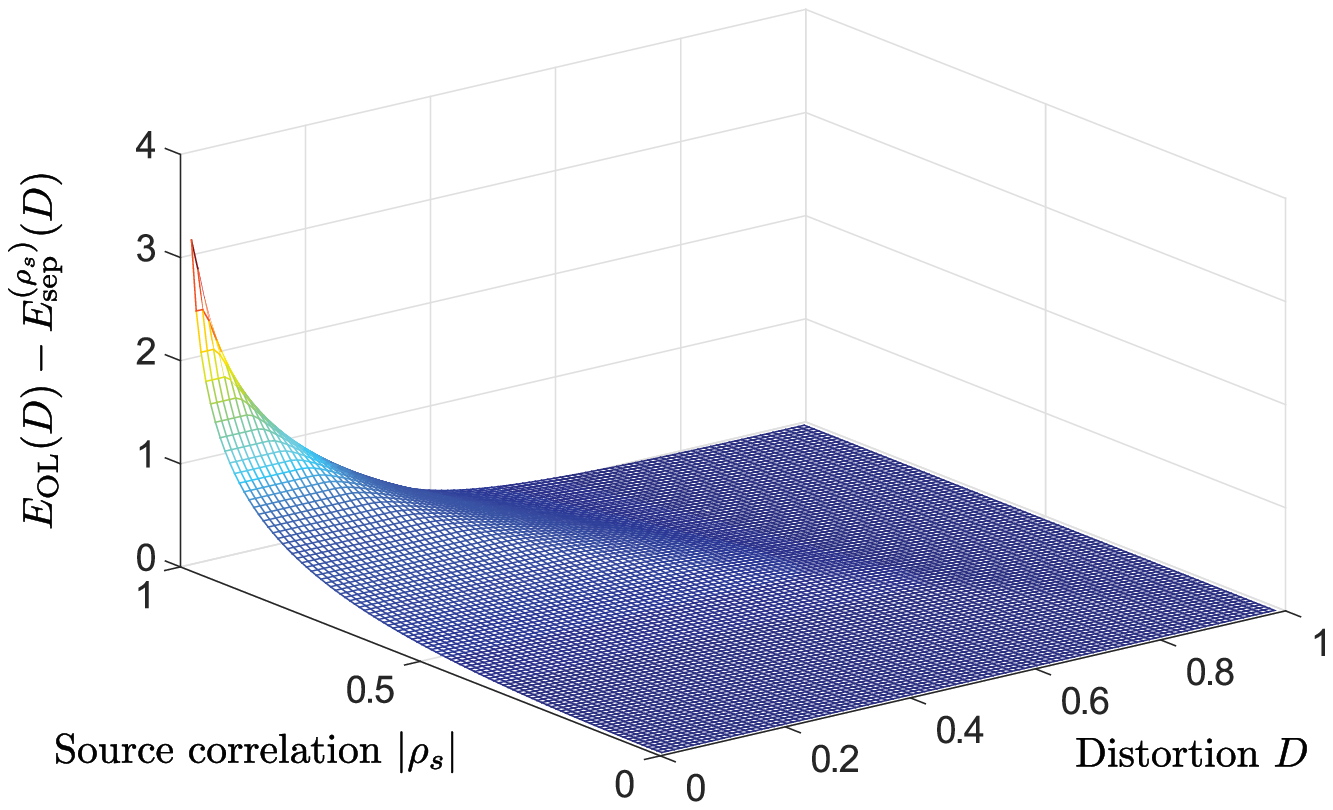}
		\vspace{-0.4cm}
		\captionof{figure}{Normalized excess energy requirement of the OL scheme over the SSCC-$\rho_s$ scheme, $\rho_z = 0.5$.}
		\label{fig:Gap_sep_ol}
		\vspace{-0.4cm}
	\end{minipage}
	\hspace{1.5cm}
	\begin{minipage}{.41\textwidth}
		\centering
		\includegraphics[width=1\linewidth]{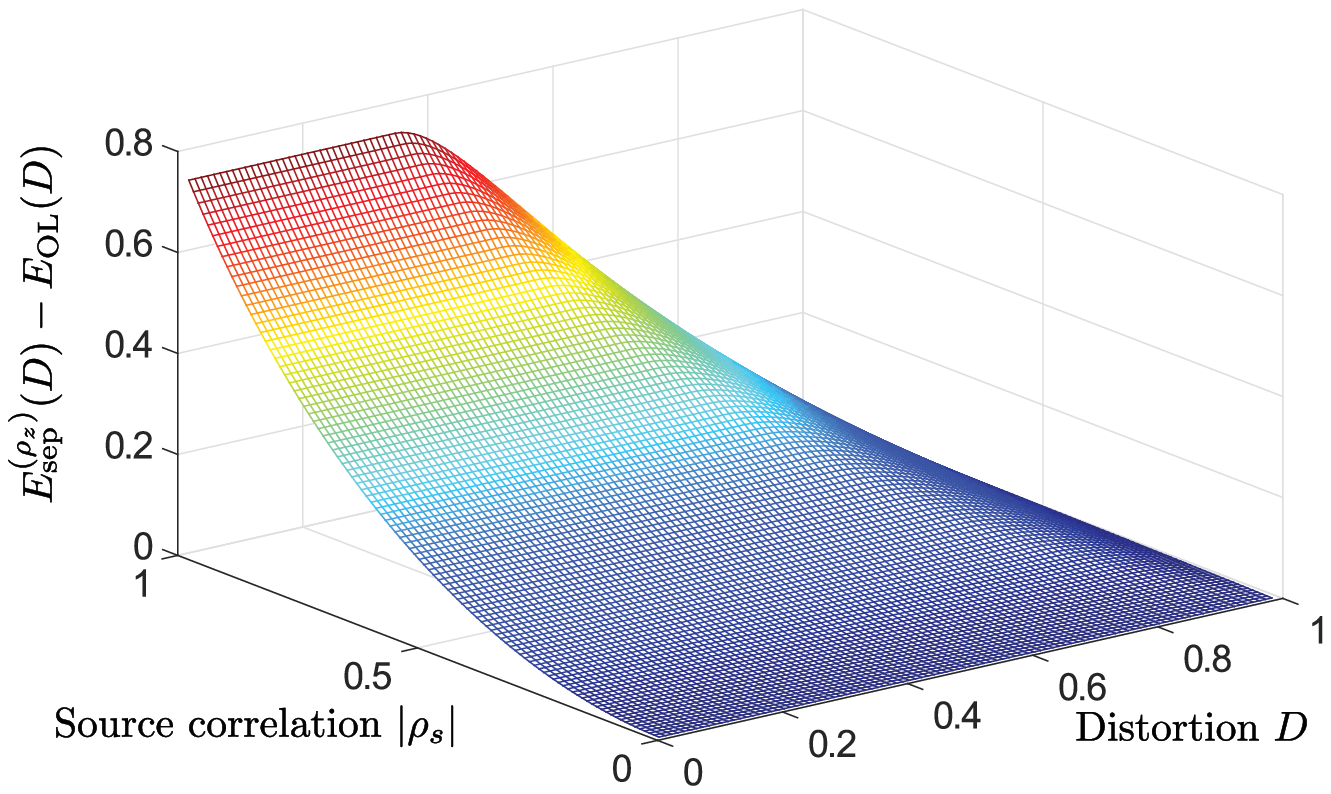}
		\vspace{-0.4cm}
		\captionof{figure}{Normalized excess energy requirement of the SSCC-$\rho_z$ scheme over the OL scheme, $\rho_z = 0.5$.}
		\label{fig:Gap_sep_lqg}
		\vspace{-0.4cm}
	\end{minipage}
\end{figure*}

\vspace{-0.05cm}
Next, we numerically compare $E_{\text{lb}}(D), E_{\text{sep}}^{(\rho_s)}(D), E_{\text{sep}}^{(\rho_z)}(D)$ and $E_{\text{OL}}(D)$. We set $\sigma_s^2 \mspace{-3mu} = \mspace{-3mu} \sigma_z^2 \mspace{-3mu} = \mspace{-3mu} 1$ and consider several values of $\rho_z$ and $\rho_s$. Fig. \ref{fig:ED_fixed_rz} depicts $E_{\text{lb}}(D), E_{\text{sep}}^{(\rho_s)}(D), E_{\text{sep}}^{(\rho_z)}(D)$ and $E_{\text{OL}}(D)$ for $\rho_z=0.5$, and for two values of $\rho_s$: $\rho_s=0.2$ and $\rho_s=0.9$. As $E_{\text{sep}}^{(\rho_z)}(D)$ is not a function of $\rho_s$, {\em it is plotted only once}.
It can be observed that when $\rho_s=0.2$, $E_{\text{sep}}^{(\rho_s)}(D), E_{\text{sep}}^{(\rho_z)}(D)$ and $E_{\text{OL}}(D)$ are almost the same. This follows because when the correlation between the sources is low, the gains from accounting for this correlation are also low. Furthermore, when $\rho_s=0.2$, the gap between the lower bound and the upper bounds is evident.
On the other hand, when $\rho_s=0.9$, both SSCC-$\rho_s$ and OL significantly improve upon SSCC-$\rho_z$. This follows as SSCC-$\rho_z$ does not take advantage of the correlation among the sources.
It can further be observed that when the distortion is low, there is a small gap between OL and SSCC-$\rho_s$, while when the distortion is high, OL and SSCC-$\rho_s$ require roughly the same amount of energy. This is also supported by Fig \ref{fig:Gap_sep_ol}.
We conclude that as the SSCC-$\rho_s$ scheme encodes over long sequences of source samples, it better exploits the correlation among the sources compared to the OL scheme.

Fig. \ref{fig:ED_fixed_rs} depicts $E_{\text{lb}}(D), E_{\text{sep}}^{(\rho_s)}(D), E_{\text{sep}}^{(\rho_z)}(D)$ and $E_{\text{OL}}(D)$ vs. $D$, for $\rho_s=0.8$, and for $\rho_z \in \{-0.9,0.9 \}$. As $E_{\text{sep}}^{(\rho_s)}(D)$ and $E_{\text{sep}}^{(\rho_z)}(D)$ are not functions of $\rho_z$, {\em we plot them only once}.
It can be observed that when $\rho_z = 0.9$, $E_{\text{lb}}(D), E_{\text{sep}}^{(\rho_s)}(D)$ and $E_{\text{OL}}(D)$ are very close to each other, as was analytically concluded in Remark \ref{rem:OL_discussion}. On the other hand, for $\rho_z = -0.9$ the gap between the bounds is large.

 Note that analytically comparing $E_{\text{sep}}^{(\rho_s)}(D)$, $E_{\text{sep}}^{(\rho_z)}(D)$ and $E_{\text{OL}}(D)$ for any $D$ is difficult. Our numerical simulations suggest that $E_{\text{sep}}^{(\rho_s)}(D) \mspace{-3mu} \le \mspace{-3mu} E_{\text{OL}}(D) \mspace{-3mu} \le \mspace{-3mu} E_{\text{sep}}^{(\rho_z)}(D)$, for all values of $D,\rho_s,\rho_z$. For example, Fig. \ref{fig:Gap_sep_ol} depicts the difference $E_{\text{OL}}(D) \mspace{-3mu} - \mspace{-3mu} E_{\text{sep}}^{(\rho_s)}(D)$ for $\rho_z = 0.5$, and for all values of $D$ and $|\rho_s|$.  
 It can be observed that for low $|\rho_s|$, or for high $D$, $E_{\text{sep}}^{(\rho_s)}(D) \approx E_{\text{OL}}(D)$. On the other hand, when $|\rho_s|$ is large and $D$ is low, then the SSCC-$\rho_s$ scheme improves upon the OL scheme. 
Fig. \ref{fig:Gap_sep_lqg} depicts the difference $E_{\text{sep}}^{(\rho_z)}(D) \mspace{-3mu} - \mspace{-3mu} E_{\text{OL}}(D)$ for $\rho_z \mspace{-3mu} = \mspace{-3mu} 0.5$. It can be observed that larger $|\rho_s|$ results in a larger gap.

\vspace{-0.25cm}
\section{Conclusions} \label{sec:conclusions}

\vspace{-0.15cm}
This work is a first step towards characterizing the optimal EDT for sending correlated Gaussian sources over GBCFs, where no restriction is placed on the signaling scheme or the source-channel bandwidth ratio. 
In particular, we first lower bounded the minimum required energy per source-pair sample using cut-set arguments, and then upper bounded it by analyzing three transmission schemes: Two different SSCC schemes, and the uncoded JSCC OL scheme. 
We concluded that while SSCC-$\rho_s$ successfully exploits the correlation between the sources by jointly encoding the source sequences into a single bit stream, SSCC-$\rho_z$ does not exploit the correlation between the noise components since the LQG channel coding scheme of \cite{AMM:12} achieves the same minimum energy per pair of bits as orthogonal transmission. This leads to the conclusion that SSCC-$\rho_s$ outperforms SSCC-$\rho_z$.
Numerical results indicate that SSCC-$\rho_s$ outperforms the OL scheme as well. On the other hand, the gap between the energy requirements of the two schemes is rather small. We note that, in the SSCC-$\rho_s$ scheme coding takes place over multiple samples of source pairs which introduces high computational complexity, large delays, and requires large amount of storage. 
On the other hand, the OL scheme applies {\em linear and uncoded transmission} to each source sample pair separately, which requires low computational complexity, short delays, and limited storage. Therefore, the OL scheme provides an attractive alternative for energy efficient transmission over GBCFs. 

\end{document}